\documentclass[12pt]{article}
\usepackage[dvips]{graphicx}
\usepackage{epsf,amssymb,hyperref}
\usepackage[english]{babel}

\def\noi{\noindent}
\def\nqq{\hspace{-2em}}

\def\barr{\left(\begin{array}}
\def\earr{\end{array}\right)}
\def\beq#1{\begin{equation}\label{#1}}
\def\eeq{\end{equation}}
\def\ber#1{\begin{eqnarray}\label{#1} &&\nqq}
\def\eer{\end{eqnarray}}

\newcommand{\bear}[1]{\begin{eqnarray}\label{#1}}
\newcommand{\ear}{\end{eqnarray}}

\renewcommand{\theequation}{\arabic{section}.\arabic{equation}}
\catcode`\@=11 \@addtoreset{equation}{section}\catcode`\@=12
\newcommand{\N}{ {\mathbb N} }
\newcommand{\R}{ {\mathbb R} }

\newcommand{\fnm}{\footnotemark}
\newcommand{\fnt}{\footnotetext}

\begin{document}

 \vspace{15pt}

 \begin{center}
 \large\bf

 On multidimensional cosmological solutions with scalar fields and 2-forms
 corresponding to rank-3 Lie algebras:  acceleration and small variation of G

 \vspace{15pt}

 \normalsize\bf
        A. A. Golubtsova\fnm[1]\fnt[1]{agolubtsova@yahoo.com}

 \vspace{7pt}

 \it \  Institute of Gravitation and Cosmology,
 Peoples' Friendship University of Russia,
 6 Miklukho-Maklaya Str.,  Moscow 117198, Russia \\

 \end{center}
 \vspace{15pt}

 \small\noi

 \begin{abstract}
By means of a simple model we investigate the possibility of an accelerated expansion of a 3-dimensional
subspace in the presence of the variation of the effective 4-dimensional constant obeying the experimental constraint.
Multidimensional cosmological solutions with $m$ 2-form fields and $l$ scalar fields are presented.
Solutions corresponding to rank-3 Lie algebras are singled out and discussed. Each of  solutions contain
two factor spaces: one-dimensional space $M_{1}$ and Ricci-flat space $M_{2}$.
A 3-dimensional subspace of  $M_{2}$ is interpreted as "our" space.
We show that there exists a time interval where accelerated expansion of our 3D space is compatible with a small
enough variation of the effective gravitational constant $G(\tau)$. This interval contains $\tau_0$
which is the point of minimum of  $G(\tau)$ (here $\tau$ is the synchronous time variable).
Special solutions with three phantom scalar fields are analyzed. It is shown that in the vicinity of the point $\tau_{0}$
the time variation of $G(\tau)$ decreases in the sequence of Lie algebras $A_{3}$, $C_{3}$ and $B_{3}$
when the solutions with asymptotically power-law  behavior of  scale-factors  for $\tau \to \infty$ are considered.
Exact solutions with asymptotically exponential accelerated expansion of 3D space are also considered.
 \end{abstract}
\newpage

\section{Introduction}

  One of the challenging problems of modern physics and cosmology is that of possible time-, location-, and
scale-dependent variations of the fundamental physical constants, in particular,  of Newton's gravitational constant $G$.
According to the observational data the variation of ${G}$ is admissible at the level of less $10^{-12}yr^{-1}$ and
there exists a necessity in further theoretical and experimental developments of this problem.
At present multidimensional cosmological models with diverse matter sources are widely used as a theoretical framework for describing
possible time variations of fundamental physical constants, e.g. the gravitational constant $G$, see  \cite{IM1}-\cite{IMA} and
references therein.

This paper is devoted to the investigation of an accelerated expansion of our 3-dimensional space in presence of the
variation of the effective 4-dimensional gravitational constant. Here we use the approach proposed
in the papers \cite{DIKM}-\cite{FI-07}. In \cite{IMK}, multidimensional exact S-brane solutions
with the intersection rules for branes corresponding to rank-2 Lie algebras were discussed.
It was shown that there exists an interval of the synchronous time $\tau$ where the scale factor
of our 3-dimensional space exhibits an accelerated expansion according to the observational data \cite{Riess,Perl}
while the relative variation of the effective 4-dimensional gravitational constant is small
enough with the Hubble parameter, see \cite{Pit}-\cite{Dic} and references therein.

In this paper, we study the variation of $G$ using a multidimensional gravitational model with
an arbitrary number of dimensions, $m$ Abelian gauge 2-form fields and $l$ scalar fields.
The 2-form fields contribute to  0-branes. Thus we have the multidimensional model with $m$ 0-branes.
Our exact solutions are governed by polynomials $H_{s}$ corresponding to rank-3 Lie algebras. We aim to show that for
this case there is an interval of the synchronous time $\tau$ where an accelerated expansion of
 ``our'' 3-dimensional space co-exists with a small enough value of $\dot G$, like in \cite{IMK}. But we also intend to show that
 the accelerated expansion is possible in the presence of "phantom" scalar fields in the model.

The paper is organized as follows. In Section 2, we consider the multidimensional gravitational model and exact solutions.
We also single out solutions corresponding to rank-3 Lie algebras in Section 2. In Section 3, the solutions are examined for the
presence of an accelerated expansion if there is a small variation of the gravitational constant. In Section 4,
a special configuration  with three phantom fields is considered and the expressions for the dimensionless variation of G corresponding
to the  rank-3 Lie algebras are present. Section 5 is devoted to the investigation of the solutions with asymptotically exponential
accelerated expansion of 3-dimensional space and Section 6 is a conclusion.

\section{The model}

 The model is governed by the action
\beq{2.1}
  S=\int d^Dx \sqrt{|g|} \biggl \{R[g]-
  h_{\alpha\beta}g^{MN}\partial_M\varphi^{\alpha}\partial_N\varphi^{\beta}-\frac{1}{2}
  \sum_{s =1}^{m}\exp[2\lambda_s(\varphi)](F^s)^2 \biggr \}
 \eeq

where $g=g_{MN}(x)dx^M\otimes dx^N$ is a D-dimensional metric of the pseudo-Euclidean signature (-,+,...,+),
$ F^s = dA^s=  \frac{1}{2} F^s_{M N}  dz^{M} \wedge  dz^{N}$
 is a $2$-form of rank 2, $\varphi=(\varphi^\alpha)\in\R^l$ is a vector of scalar fields,
 $(h_{\alpha\beta})$ is a constant symmetric non-degenerate
 $l\times l$ matrix $(l\in \N)$, $\lambda_s$ is a 1-form on $\R^l$:
 $\lambda_s(\varphi)=\lambda_{s \alpha}\varphi^\alpha$, with
 $s = 1,..., m$ and $\alpha=1,\dots,l$.
 In (\ref{2.1})
 we denote $|g| =   |\det (g_{MN})|$, $(F^s)^2  =
 F^s_{M_1 M_{2}} F^s_{N_1 N_{2}}  g^{M_1 N_1} g^{M_{2} N_{2}}$, $s = 1,..., m$.

We consider the manifold
\beq{2.2}
  M = (0, + \infty)  \times M_1 \times M_2,
 \eeq
 where $M_1$ is a one-dimensional manifold (say $S^1$ or $\R$) and
 $M_2$ is a (D-2)-dimensional Ricci-flat manifold.\\
\subsection{General solutions.}
  In what follows the subspace $M_{1}$ will support all forms $A^{s}$.
  Let us consider a family of exact
solutions to the field equations corresponding to the action
(\ref{2.1}) and depending on one variable $\rho$. These solutions
  are defined on the manifold (\ref{2.2}).
The solutions read \cite{GIGr}
 \bear{3.30}
  g= \Bigl(\prod_{s = 1}^{m} H_s^{2 h_s /(D-2)} \Bigr)
  \biggl\{ w d\rho \otimes d \rho  +
  \Bigl(\prod_{s = 1}^{m} H_s^{-2 h_s} \Bigr) \rho^2 d\phi \otimes d\phi +
    g^2  \biggr\},
 \\  \label{3.31}
  \exp(\varphi^\alpha)=
  \prod_{s = 1}^{m} H_s^{h_s  \lambda_{s}^\alpha},
 \\  \label{3.32a}
  F^s= - Q_s \left( \prod_{s' = 1}^{m}  H_{s'}^{- A_{s
  s'}} \right) \rho d\rho \wedge d \phi,
  \ear
 $s = 1,..., m$, where $w = \pm 1$; $g^1 = d\phi \otimes d\phi$ is a
  metric on $M_1$ and $g^2$ is a Ricci-flat metric on
 $M_{2}$.

 Functions $H_s(z) > 0$ with $z = \rho^2$, are defined on the interval $(0, +\infty)$ and obey the
  non linear differential equations   \beq{1.1}
  \frac{d}{dz} \left( \frac{ z}{H_s} \frac{d}{dz} H_s \right) =
   P_s \prod_{s' = 1}^{m}  H_{s'}^{- A_{s s'}},
  \eeq
 with  the following boundary conditions imposed
 \beq{1.2}
   H_{s}(+ 0) = 1.
 \eeq

 Parameters  $P_s$ are proportional to the charge density squared parameters in the following way:
  \beq{2.21}
  P_s =  \frac{1}{4} K_s Q_s^2.
 \eeq
 Parameters  $h_s$  satisfy the relations
  \beq{2.16}
  h_s = K_s^{-1}, \qquad  K_{s} = (U^{s},U^{s}),
  \eeq
 where $K_{s} \neq 0$ and the scalar products of the $U^{s}$-vectors belonging to $\R^{n + l}$ are defined
 as follows
 \beq{2.17}
  (U^{s},U^{s'}) =
  1 +\frac{1}{2-D} +  \lambda_{s \alpha} \lambda_{s' \beta}   h^{\alpha\beta},
  \eeq
 $s = 1,..., m$, with $(h^{\alpha\beta})=(h_{\alpha\beta})^{-1}$ and
 $\lambda_{s}^{\alpha} = h^{\alpha\beta}  \lambda_{s \beta}$. The $U^{s}$-vectors and the scalar
 products were specified in \cite{IMmul,IMpb,IMsig}.

 Here we put the matrix
 \beq{2.18}
  (A_{ss'}) = \left( 2 (U^{s},U^{s'})/(U^{s'},U^{s'}) \right)
 \eeq
  to be coinciding with the Cartan matrix for a simple Lie algebra $\cal G$ of rank $m$.
  If we remember the integrality condition for a root system and compare it with (\ref{2.18}), we will notice
  a close correspondence  between $U^{s}$-vectors and roots.

  According to a conjecture  suggested in \cite{Iflux, GIM}
  solutions to eqs. (\ref{1.1}), (\ref{1.2})
 (governed by the Cartan matrix $(A_{s s'})$)
 are  polynomials:
 \bear{Hs}
 H_{s}(z) = 1 + \sum_{k = 1}^{n_s} P_s^{(k)} z^k,
 \ear
  where
 \bear{Vns}
 n_s = 2 \sum_{s' =1}^{n} A^{s s'}
 \ear and  $(A^{s s'}) = (A_{s s'})^{-1}$.
 The integers $n_s$ are components  of the  twice  dual
 Weyl vector in the basis of simple co-roots \cite{GIM, FS}.
 It should be also noted that the set of polynomials $H_{s}$ defines a special solution to the open Toda
 chain equations corresponding to a simple Lie algebra $\cal G$ \cite{GI}.

 These solutions are special cases of more general solutions from \cite{Iflux}.
 The solutions under consideration may be verified just by the substitution
 into the equations of motion corresponding to
 (\ref{2.1}). It may be also obtained as a special case of the
 fluxbrane (for $w = +1$,  $M_1 = S^1$) and $S$-brane
 ($w = -1$) solutions from \cite{Iflux} and \cite{GIM}, respectively.\\

 \subsection{Solutions for Lie algebras of rank 3.}

 Let us single out a special class of exact solutions corresponding to rank-3 Lie algebras $A_{3}$, $B_{3}$ and $C_{3}$.

 The Cartan matrix corresponding to Lie algebras rank 3 reads
\bear{ABC}
 A = \left(
   \begin{array}{ccc}
     2 & -1 & 0 \\
     -1 & 2 & -k_{1} \\
     0 & -k_{2} & 2 \\
   \end{array}
 \right),
 \ear
where here and in the sequel
 \bear{k1k2}
 k_{1}=(1,2,1), \quad k_{2} = (1,1,2),
 \ear
for the Lie algebras $A_{3}$, $B_{3}$ and $C_{3}$, respectively.

 Due to (\ref{2.18}) and (\ref{ABC}), we get
 \bear{K}
 K_{1} = K_{2} =\displaystyle{\frac{k_{1}}{k_{2}}K_{3}},
 \ear
 denoting $K_{1}=K_{2} = K$ and $K_{3} = K'$, we get $K=\displaystyle{\frac{k_{1}}{k_{2}}K'}$.

 Here we put $K < 0$ and use a special choice of $P_{s}$-parameters:
  \bear{parP}
  P_{s} = n_{s}P,
  \ear
  $P \neq 0$,
  where the integers $n_{s}$ are components of a twice dual Weyl vector in the basis of simple co-roots:
  \bear{n123}
   (n_{1}, n_{2}, n_{3}) = (3,4,3),(6,10,6),(5,8,9)
  \ear
   for the Lie algebras $A_{3}$, $B_{3}$ and $C_{3}$ respectively. Then due to relations
  (\ref{2.21}) and (\ref{parP}) we obtain
  \bear{parPQ}
     Q^{2}_{s} = \frac{4n_{s}}{K_{s}}P.
     \ear

   The governing functions in this case have the following form
  \bear{BINN}
  H_{s} = (1 + P t^{2})^{n_{s}} = X^{n_{s}},
  \ear
 where
  \bear{funcH}
  X = 1 + P t^{2}
  \ear
  and $t$ is time variable.

  For general form of polynomials $H_{s}$ corresponding to the Lie algebras $A_{3}$, $B_{3}$, $C_{3}$
  see Appendix A.

  The solutions read
  \bear{gA3}
   g = X^{2A} \biggl\{ - dt \otimes dt + X^{-2 B} t^2 d\phi \otimes d\phi
   + g^{2}  \biggr\},
   \\ \label{phA3}
  \exp(\varphi^\alpha) = X^{B_{1}\lambda_{1}^\alpha + B_{2}\lambda_{2}^\alpha + B_{3}\lambda_{3}^\alpha},
  \\ \label{FA3}
   F^{1} = - Q_{1}  X^{n_{2} - 2n_{1}} t dt \wedge d \phi,
   \\ \label{FA32}
   F^{2} = - Q_{2}  X^{n_{1} - 2n_{2}+k_{1}n_{3}} t dt \wedge d \phi,
   \\ \label{FA33}
   F^{3} = - Q_{3}  X^{k_{2}n_{2} - 2n_{3}} t dt \wedge d \phi,
  \ear
  where
  \bear{parAB}
   A = \frac{B}{D-2},
   \\ \label{parBs}
   B_{s} = n_{s} K^{-1}_{s}, \quad B = \sum^{3}_{s=1}B_{s},
  \ear
  $s = 1,2,3$.

 The relations (\ref{FA3})-(\ref{FA33}) mean that the only nonzero components of the electromagnetic field tensor
 are $F^{s}_{ty} = - F^{s}_{yt} = -Q_{s}X^{-(n_{1}A_{s1} + n_{2}A_{s2}+ n_{3}A_{s3})}t$.

 We also note that the charge density parameters $Q_{s}$ obey the following relations (see (\ref{K}) and (\ref{parPQ}))
 \bear{Q12}
 \frac{Q_{1}^{2}}{Q_{2}^{2}} = \frac{n_{1}}{n_{2}}\frac{K_{2}}{K_{1}} = \left(\frac34, \frac35, \frac58\right), \quad
 \frac{Q_{2}^{2}}{Q_{3}^{2}} = \frac{n_{2}}{n_{3}}\frac{k_{2}}{k_{1}} = \left( \frac43, \frac56, \frac{16}{9}\right)
 \ear
for the Lie algebras $A_{3}$, $B_{3}$ and $C_{3}$, respectively.

\section{\bf Solutions with acceleration}

 In what follows we use a "synchronous" time variable $\tau =\tau(t)$ :
  \bear{Tau}
  \tau = \int^{t}_{0}d\bar{t}[X(\bar{t})]^{A}.
  \ear

  Recall that  $P < 0$,  $K < 0$ and hence $A < 0$.
  Consider two intervals of parameter $A$:
 \bear{2.2i}
    {\bf (i)} \ \ \ A  < -1,
 \\  \label{2.2ii}
    {\bf (ii)}  \ \ -1 < A  < 0.
 \ear
 (A special case $A= -1$ will be investigated in other section.)

 In both cases after integrating we get Gauss's hypergeometric
 functions. If we look at the graphics of these functions (Fig. 1 and Fig. 2),
 we note that in case (i), the function $\tau = \tau (t)$
monotonically increases from $0$ to $+ \infty$, for $t \in (0, t_1)$, where $t_1
 = |P|^{-1/2}$, while in case $(ii)$ it is monotonically increases
from 0 to a finite value $\tau_1 = \tau(t_1)$.
\begin{figure}[ht] \centering
 \parbox[b]{0.49\textwidth}{\centering
 \includegraphics[width=6.3 cm,height=5 cm]{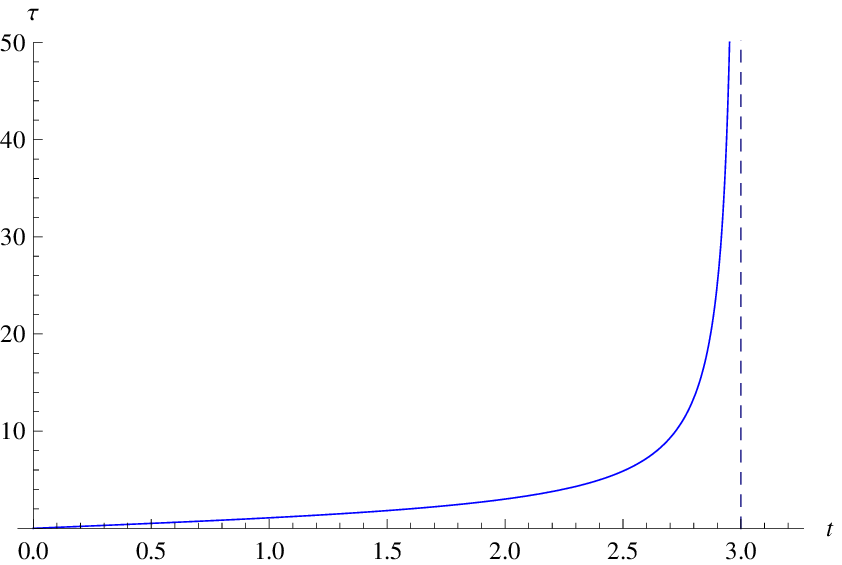}
 \caption{The graph of the "synchronous" time variable $\tau$, for $A < -1$}}
\hfil \hfil \hfil
\parbox[b]{0.49\textwidth}{\centering
 \includegraphics[width=6.3 cm,height=5 cm]{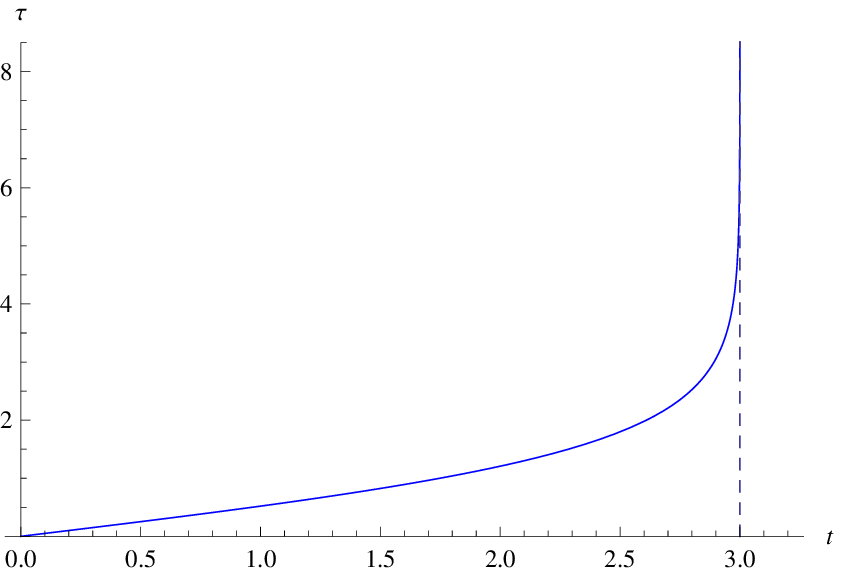}
 \caption{The graph of the "synchronous" time variable $\tau$, for $-1 < A < 0$}}
\end{figure}

The solutions  (\ref{gA3})-(\ref{FA33}) are defined on the manifold
\bear{Man3}
M = (0,t_{1}) \times M_{1} \times M_{2}.
\ear
Here we put $M_{1} = S^{1}$ and
\bear{Man32}
M_{2} = R^{3} \times (S^{1})^{n-4},
\ear
where the subpace $\R^{3}$ is our 3-dimensional space with the scale factor
 \bear{scf}
 a_{3} = X^{A}.
 \ear

 For the first branch $(i)$ we get an asymptotical relation
 \bear{ai}
 a_{3} \sim  {\rm const }   \tau^{\nu},
 \ear
 for $\tau \rightarrow + \infty$, where
 \bear{nu}
 \nu = \frac{A}{A + 1}
 \ear
 and, due to (\ref{2.2i}), $\nu > 1$. For the second branch $(ii)$ we obtain
 \bear{aii}
 a_{3} \sim {\rm const}  (\tau_{1} - \tau)^{\nu},
 \ear
 for $\tau \rightarrow \tau_{1} - 0$, where $\nu < 0$, due to $-1 < A < 0$.

 Thus, we get an asymptotic accelerated expansion of the 3D subspace $\R^{3}$ in both
 cases $(i)$ and $(ii)$, and $a_{3} \rightarrow +\infty$.

 It may be verified that the accelerated expansion takes place for
 all $\tau > 0$, i.e.,
 \bear{dda}
 \dot{a}_{3} > 0,\qquad \ddot{a}_{3} > 0
 \ear
  (see \cite{IMK,IMA}).

 Figure 3 and Figure 4 show a behavior of the function $a_{3}(t)$  and its asymptotic forms.

\begin{figure}[ht] \centering
 \parbox[b]{0.49\textwidth}{\centering
 \includegraphics[width=6.3 cm,height=5 cm]{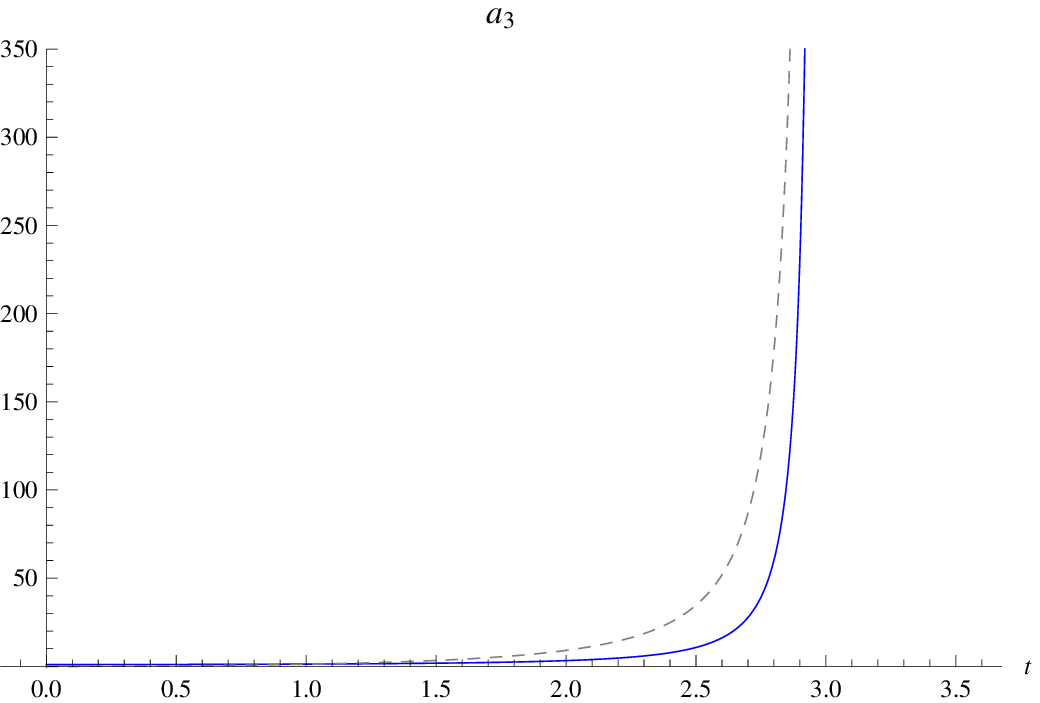}
 \caption{The scale factor $a_{3}(t)$ and its asymptotic form, for $A < -1$}}
\hfil \hfil \hfil
\parbox[b]{0.49\textwidth}{\centering
 \includegraphics[width=6.3 cm,height=5 cm]{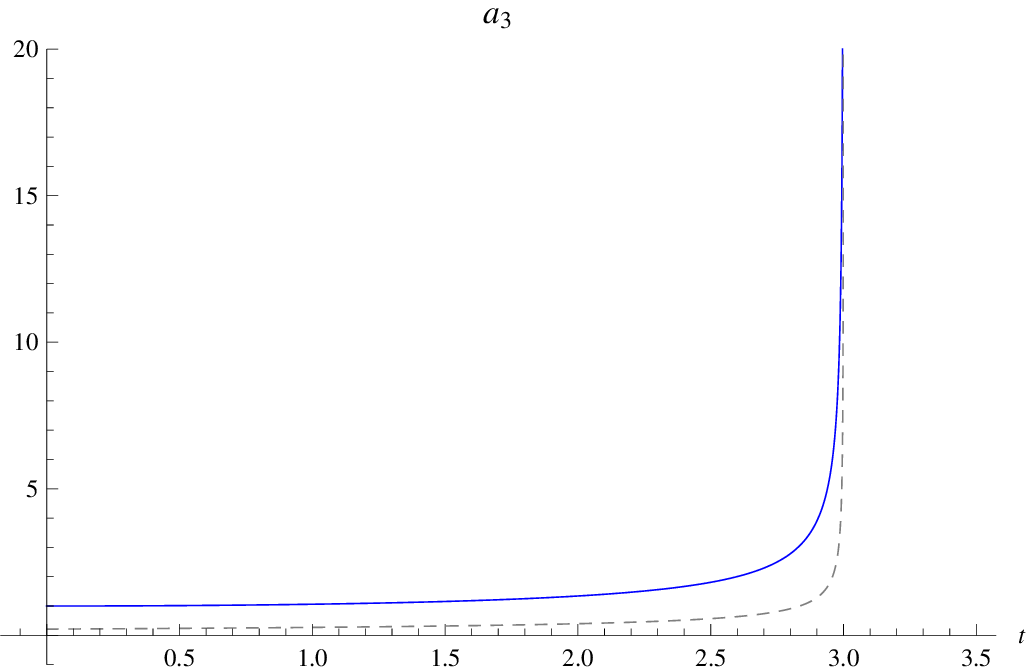}
 \caption{The scale factor $a_{3}(t)$ and its asymptotic form, for $-1 < A < 0$}}
\end{figure}

 Now we consider a variation of the effective constant $G$. According to (\ref{Man3})-(\ref{Man32})
 in our model we have two internal compact spaces $S^{1}$ and $(S^{1})^{n-4}$ with scale factors
 \bear{scfac}
    a_{1} = X^{A-B}t, \qquad a_{2} = X^{A},
 \ear
 respectively.
In Jordan's frame the 4-dimensional gravitational "constant" is
  \bear{constG}
  G =  {\rm const } a_{1}^{-d_{1}} a_{2}^{-d_{2}} = {\rm const }X^{B - A(n-3)}t^{-1},
  \ear
  where $d_{1} =1$  and $d_{2} = n - 4$ are dimensions of factor spaces $S^{1}$ and $(S^{1})^{n-4}$, respectively.
Due to the formulas (\ref{parAB})-(\ref{parBs}) for $A$ and $B$ parameters and remembering that $D = n + 1$
 we rewrite the expression (\ref{constG}) as follows
  \bear{conG}
  G = X^{2A}t^{-1}.
  \ear
 The dimensionless variation of $G$ reads
 \bear {varGd}
 \delta = \dot{G}/(GH) = 2 + \frac{1 - |P|t^{2}}{2A|P|t^{2}},
 \ear
 where
 \bear{hubbl}
 H = \frac{\dot{a}_{3}}{a_{3}}
 \ear
 is the Hubble parameter of the 3D subspace  $\R^{3}$. Using the expression (\ref{varGd}) we can find an extremum of the function
 $G(\tau)$ at the point $\tau_{0}$ corresponding to $t_{0}$ and
 \bear{extr}
 t^{2}_{0} = \frac{|P|^{-1}}{1+4|A|}.
 \ear
 At this point the function $G(\tau)$ has a minimum and $\dot{G}$ vanishes thereafter.

 The function $G({\tau})$ monotonically decreases from $+ \infty$ to $G_0
 = G(\tau_0)$ for $\tau \in (0, \tau_0)$ and  monotonically increases from
 $G_0$ to $+ \infty$ for $\tau \in (\tau_0, \bar{\tau}_1)$. Here
 $\bar{\tau}_1 = +\infty $ for the case (i) and  $\bar{\tau}_1 = \tau_1$ for
the case (ii) (see Fig. 5 and Fig. 6).

\begin{figure}[ht] \centering
 \parbox[b]{0.49\textwidth}{\centering
 \includegraphics[width=6.3 cm,height=5 cm]{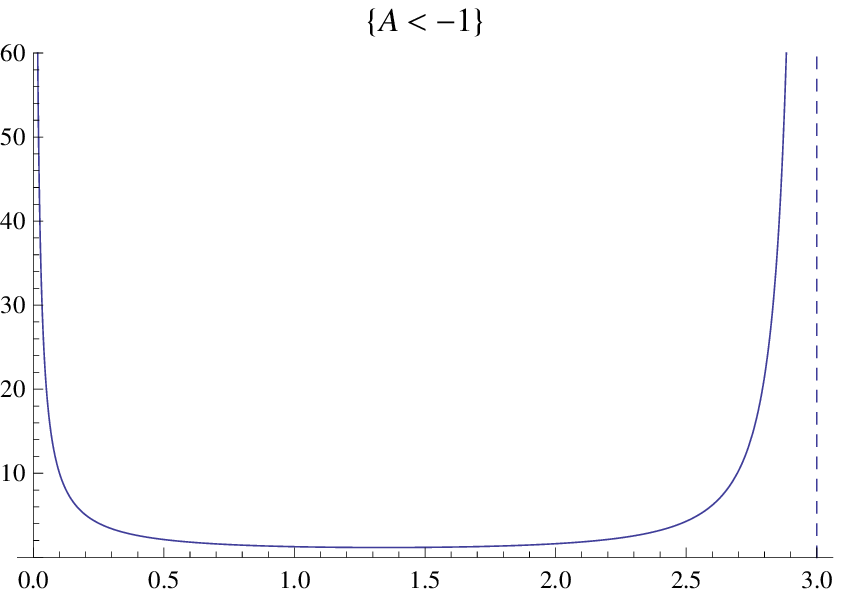}
 \caption{The plot of the gravitation constant $G(t)$, for $A < -1$}}
\hfil \hfil
\parbox[b]{0.49\textwidth}{\centering
 \includegraphics[width=6.3 cm,height=5 cm]{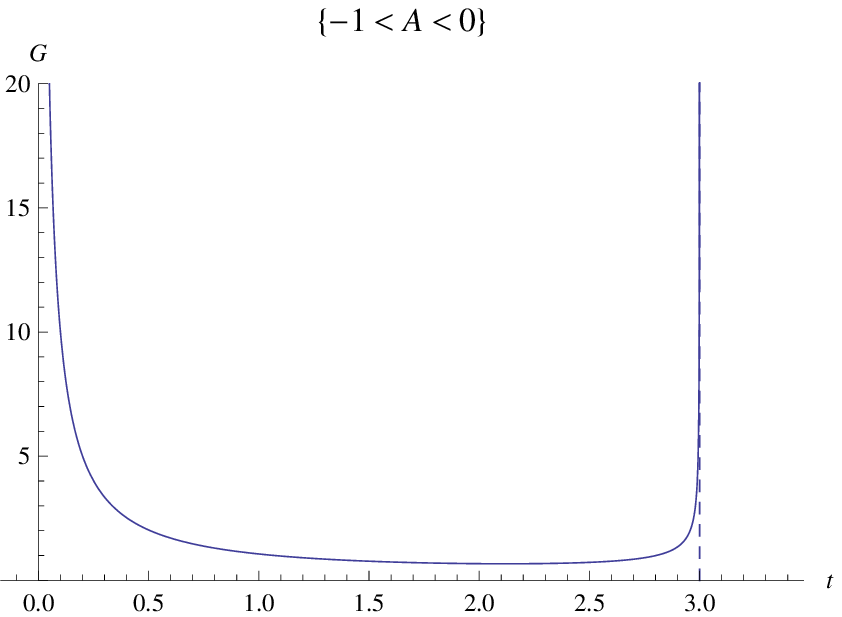}
 \caption{The plot of the gravitation constant $G(t)$, for $-1 < A < 0$}}
\end{figure}

 We should consider only solutions with the accelerated expansion of our space and small enough variations
 of the gravitational constant obeying the experimental constraint \cite{Pit}-\cite{Dic}
 \bear{delt}
 |\delta| < 0.001.
 \ear
 Here, as in the model with two form fields and two scalar fields \cite{IMK}, $\tau$ is restricted
to a certain range containing $\tau_0$. It follows from (\ref{varGd}) that in
the asymptotical regions (\ref{2.2i}) and (\ref{2.2ii}) $\delta \to 2$,
which is unacceptable due to the experimental bounds (\ref{delt}).  This
restriction is satisfied for a range containing the point $\tau_0$ where
$\delta = 0$.

Calculating  $\dot G$, in the linear approximation near $\tau_0$, we get
the following approximate relation for the dimensionless parameter
of relative variation of $G$:
 \beq{Appro}
         \delta  \approx  (8 + 2 |A|^{-1}) H_0 (\tau - \tau_0),
 \eeq
where $H_0 = H(\tau_0)$. This relation  gives approximate bounds for values
of the time variable $\tau$ allowed by the restriction on $\dot
 G$. A derivation of this relation is given in Appendix B.

The solutions under consideration with $P < 0$, one-dimensional $M_{1}$ and $3D$ subspace
 $\R^{3}$  take place when the configuration of 2-form fields, the matrix
 $(h_{\alpha\beta})$ and the dilatonic coupling vectors
 $\lambda_a$, obey the relations (\ref{2.16}), (\ref{2.17}) and
 $K_s < 0$. This is  possible when $(h_{\alpha\beta})$ is not
positive-definite, otherwise  all $K_s > 0$. Thus, there should be
at least one scalar field with a negative kinetic term (i.e.,
 a phantom scalar field).

 \section{\bf Example: a model with three phantom fields}

 Let us consider the following example of the cosmological solutions: $l=3$, $(h_{\alpha \beta}) = - (\delta_{\alpha \beta})$, $w = -1$,
 i.e. there are three phantom scalar fields.

 Using (\ref{2.16}), (\ref{2.17}) and (\ref{K}) we get that the coupling vectors obey the following relations:
 \bear{lamb12}
 \lambda^{2} = \vec{\lambda}^{2}_{1} = \vec{\lambda}^{2}_{2} = 1 + \frac{1}{2 - D} - K, \qquad
 \vec{\lambda}^{2}_{3}  = 1 + \frac{1}{2 - D} - K',
 \ear
 \bear{lamb13}
 \vec{\lambda}_{1}\vec{\lambda}_{2} =  1 + \frac{1}{2 - D} + \frac12 K, \quad
 \vec{\lambda}_{2}\vec{\lambda}_{3} =  1 + \frac{1}{2 - D} + \frac{k_{2}}{2} K, \quad
 \vec{\lambda}_{1}\vec{\lambda}_{3} =  1 + \frac{1}{2 - D},
 \ear
 where $K < 0$ and $\displaystyle {K' = \frac{k_{2}}{k_{1}}K}$, $k_{1} = 1,2,1$, $k_{2} = 1,1,2$ for
 algebras $A_{3}$, $B_{3}$ and $C_{3}$, respectively.
 It was verified (i.e. by the use of Mathematica) that the matrix of scalar
  products
  \bear{Sclambd}
   (\vec{\lambda}_{s}\vec{\lambda}_{s'})
             = \left(
                    \begin{array}{ccc} \displaystyle{
                      M - K} & \displaystyle{ M +
                      \frac12 K} & \displaystyle{M} \\
                       \displaystyle{M +
                       \frac12 K} &\displaystyle{ M - K}
                        & \displaystyle{M + \frac{k_{2}}{2} K} \\
                      \displaystyle{M} & \displaystyle{ M + \frac{k_{2}}{2}K}
                      & \displaystyle{M - \frac{k_{2}}{k_{1}}K} \\
                    \end{array}
                  \right),
  \ear
  with $M = 1 + \displaystyle{\frac{1}{2-D}} > 0$ ($D \geq 5$), is positive definite  for all $K < 0$
   and hence the set of vectors obeying
   (\ref{lamb12}), (\ref{lamb13})  does exist. Thereby $(\vec{\lambda}_{s}\vec{\lambda}_{s'})$ is the Gramian matrix.

 Now we compare the $A$ parameters corresponding to different Lie algebras $A_{3}$, $B_{3}$ and $C_{3}$, when the parameter $K$ is fixed.
 We get from the definition (\ref{parAB})
 \bear{parAk}
 A = \frac{1}{K (D-2)}\left(n_{1} + n_{2} + n_{3}\frac{k_{1}}{k_{2}}\right),
 \ear
 or, in detail , (see (\ref{n123}))
  \bear{3.5a}
      A_{(1)} = \frac{10}{K (D-2)},
    \qquad
     A_{(2)} = \frac{28}{K (D-2)},
    \qquad
     A_{(3)} = \frac{35}{2K (D-2)},
  \ear
  for Lie algebras $A_{3}$, $B_{3}$ and $C_{3}$,  respectively.
  Hence,
    \beq{3.6}
        |A_{(1)}| < |A_{(3)}| < |A_{(2)}|.
   \eeq

   Due to the relation (\ref{Appro}) for the dimensionless parameter of
  the relative variation of $G$ calculating in the leading approximation
   when $(\tau - \tau_0)$ is small, we get for approximate values
   of $\delta$: $\delta_{(1)}^{ap} > \delta_{(3)}^{ap} >
   \delta_{(2)}^{ap}$ that means that the variation of $G$
   (calculated near $\tau_0$)   decreases  in the sequence
   of Lie algebras $A_{3}$, $C_{3}$ and $B_{3}$, but
   the allowed interval $\triangle \tau = \tau - \tau_0$ (
    obeying $|\delta| < 0.001$)  increases in the sequence
    of Lie algebras $A_{3}$, $C_{3}$ and $B_{3}$.
   This effect could be strengthen (even drastically) when $|K_1|$ becomes
   larger.   We note that for $|K| \to + \infty$ we get a strong
   coupling limit   $\vec{\lambda}_a^2 \to + \infty$, $a = 1,2,3$,
   due to the relations (\ref{lamb12}).

 \section{Solutions with asymptotically exponential acceleration}

  We deal here with a special case of the solutions (\ref{gA3}) - (\ref{FA3}) when the parameter $A = -1$.
  As in the previous section we have three scalar "phantom" fields and the dimension is arbitrary.

 In this case the relations for the vector couplings (\ref{lamb12})-(\ref{lamb13}) remain unchanged.
 Due to $A = -1$ the synchronous time variable $\tau = \tau(t)$ is defined by
 the relation:
  \beq{tauA1}
  \tau = \int_{0}^{t} d \bar{t} [X(\bar{t})]^{-1} = \frac{Arctanh(\sqrt{|P|}t)}{\sqrt{|P|}}.
 \eeq
  The function $\tau = \tau(t)$ is monotonically increasing from $0$ to $+ \infty$ for $t \in (0,t_{1})$, where $t_{1} = |P|^{-1/2}$.

  The solutions are defined on the manifold
 \bear{ManR}
  M = (0, + \infty)\times M_{1} \times M_{2},
 \ear
 and $M_{2}$  is a (D-2)-dimensional Ricci-flat factor-space.
 Using the variable $\tau$ we can write the solutions in terms of the synchronous time variable:
   \bear{gA1}
   g = - d\tau \otimes d\tau + \bar{X}^{2(D-3)} Y^{2} dy \otimes dy + \bar{X}^{-2} g^{2} ,
   \\ \label{phA1}
  \varphi^{\alpha} = - h \left(n_{1}\lambda_{1\alpha} + n_{2}\lambda_{2\alpha} + n_{3}\lambda_{3\alpha}\frac{k_{1}}{k_{2}}\right)\ln{\bar{X}},
  \\ \label{FA1}
   F^{1} = - Q_{1}  \bar{X}^{n_{2} - 2n_{1} + 1} Y d\tau \wedge dy,
   \\ \label{FA12}
   F^{2} = - Q_{2}  \bar{X}^{n_{1} - 2n_{2}+k_{1}n_{3} + 1} Y d\tau \wedge dy,
   \\ \label{FA13}
   F^{3} = - Q_{3} \bar{X}^{k_{2}n_{2} - 2n_{3} + 1} Y d\tau \wedge dy,
  \ear
 where
  \bear{parh1}
   \bar{X} = \frac{1} {\cosh^{2}{\sqrt{|P|}\tau}}, \qquad  Y = \frac{\tanh(\sqrt{|P|}\tau)}{\sqrt{|P|}}, \\
   h = \frac{2 - D}{3 - D - \lambda^{2}(2 - D)}
  \ear
 the charge density parameters $Q_{s}$ satisfy (\ref{parPQ}) , $s = 1,2,3$.
 Due to (\ref{3.5a}) $K(D - 2) = (-10, -28, -35/2)$ for Lie algebras $A_{3}$ , $B_{3}$ and $C_{3}$, respectively.

  In the factor-space $M_{2} = \R^{3}\times(S^{1})^{n-4}$ we single out "our" three-dimensional space with
  the scale factor
  \bear{sca1}
  a_{3} = \bar{X}^{-1}.
  \ear
  and when $\tau \rightarrow + \infty$ we get the asymptotic formula
 \bear{assiA1}
  a_{3} \sim \frac{1}{4}\exp (2\tau/t_{1}).
 \ear
 Hence our three-dimensional space expands exponentially and $a\rightarrow + \infty$ as $\tau \rightarrow +\infty$.

 Now, let us consider the variation of $G$.  Using the mechanism examined earlier
 we get the approximate relation for dimensionless variation of $G$
 \bear{delA1}
 \delta \approx 10 H_{0} (\tau -\tau_{0}).
 \ear

 \section{Conclusions}

 We have considered a family of exact cosmological solutions
 in the multidimensional model with scalar and Abelian gauge fields.
 We have singled out the solutions corresponding to rank-3 Lie algebras.

 Here,  as for electric S-brane solutions \cite{IMK},
 we have found that there exists a time interval where
accelerated expansion of our 3-dimensional space is compatible
with a small enough value of $\dot{G}/G$ obeying the experimental
bounds. This interval contains  a point of minimum of the function
$G(\tau)$ denoted as $\tau_0$. It was shown there should be at least
one scalar field with negative kinetic term to ensure an accelerated
expansion of 3D space.

We have analyzed the special  solutions with three phantom scalar fields
for the Lie algebras $A_{3}$,$B_{3}$,$C_{3}$.  In the vicinity of the point  $\tau_0$
 the time variation of $G(\tau)$ (calculated in the linear approximation)
 decreases  in the sequence of Lie algebras $A_{3}$, $C_{3}$ and $B_{3}$. A generalization of
 this result to the case of $m$-forms will be a subject of a separate publication.

Among these solutions an example of solutions with exponential dependence of the scale factor
$a_{3}$ (w.r.t. sinchronous time variable) was presented.

\renewcommand{\theequation}{\Alph{subsection}.\arabic{equation}}
\renewcommand{\thesection}{}
\renewcommand{\thesubsection}{\Alph{subsection}}
\setcounter{section}{0}

\section{Appendix}

\subsection{General form of polynomials}

{\bf $A_3$-case.} The polynomials for the $A_3$-case read as
 follows

 \bear{A.8}
 H_{1} = 1 + P_1 z + \frac14 P_1 P_2 z^{2} + \frac{1}{36} P_1 P_2 P_3 z^{3},\\
 \label{A.9}
 H_{2} = 1 + P_2 z + \Bigl( \frac14 P_1 P_2 + \frac14 P_2 P_3 \Bigr) z^{2} + \frac19 P_1 P_2 P_3 z^{3}
  \\ \nonumber
  + \frac{1}{144} P_1 P_2^{2} P_3 z^{4},\\
 \label{A.10}
 H_{3} = 1 + P_3 z + \frac14 P_2 P_3 z^{2} + \frac{1}{36} P_1 P_2 P_3
 z^{3}.
 \ear
  \vspace{15pt}
{\bf$B_3$-polynomials.}

 For the Lie algebra $B_3$ we get the following polynomials

\bear{B.1}
H_{1} = 1 + P_1 z + \frac14 P_1 P_2 z^{2} + \frac{1}{18} P_1 P_2 P_3 z^{3}
 + \frac{1}{144} P_1 P_2 P_3^{2} z^{4} + \frac{1}{3600} P_1 P_2^{2} P_3^{2} z^{5}
   \\ \nonumber
 + \frac{1}{129600} P_1^{2} P_2^{2} P_3^{2} z^{6},\\
 \label{B.2}
 H_{2} = 1 + P_2 z + \Bigl( \frac14 P_1 P_2 + \frac12 P_2 P_3 \Bigr) z^{2}
  + \Bigl( \frac19 P_2 P_3^{2} + \frac29 P_1 P_2 P_3 \Bigr) z^{3}
   + \Bigl( \frac{1}{144} P_2^{2} P_3^{2}
  \\
\nonumber
  + \frac{1}{72} P_1 P_2^{2} P_3 + \frac{1}{16} P_1 P_2 P_3^{2} \Bigr) z^{4}
   + \frac{7}{600} P_1 P_2^{2} P_3^{2} z^{5} +
   \Bigl( \frac{1}{1600} P_1 P_2^{3} P_3^{2} + \frac{1}{5184} P_1^{2} P_2^{2} P_3^{2}
  \\ \nonumber
   + \frac{1}{2592} P_1 P_2^{2} P_3^{3} ) z^{6} + \Bigl( \frac{1}{16200} P_1 P_2^{3} P_3^{3}
    + \frac{1}{32400} P_1^{2} P_2^{3} P_3^{2} \Bigr) z^{7} + \Bigl( \frac{1}{518400} P_1 P_2^{3} P_3^{4}
   \\ \nonumber
   + \frac{1}{259200} P_1^{2} P_2^{3} P_3^{3} \Bigr) z^{8} +
   \frac{1}{4665600} P_1^{2} P_2^{3} P_3^{4} z^{9} + \frac{1}{466560000} P_1^{2} P_2^{4} P_3^{4} z^{10},\\
   \label{B.3}
   H_{3} = 1 + P_3 z + \frac14 P_2 P_3 z^{2} + \Bigl( \frac{1}{36}P_1 P_2 P_3 + \frac{1}{36}
    P_2 P_3^{2} \Bigr) z^{3} + \frac{1}{144} P_1 P_2 P_3^{2} z^{4}
   \\ \nonumber
   + \frac{1}{3600} P_1 P_2^{2} P_3^{2} z^{5} + \frac{1}{129600} P_1 P_2^{2} P_3^{3}
   z^{6}.
\ear
{\bf$C_3$-polynomials.}

The $C_{3}$ -polynomials have the  following form

\bear{C.1}
H_{1} = 1 + P_1 z + \frac14 P_1 P_2 z^{2} + \frac{1}{36} P_1 P_2 P_3 z^{3} + \frac{1}{576} P_1 P_2^{2} P_3 z^{4}
 + \frac{1}{14400} P_1^{2} P_2^{2} P_3 z^{5}, \\
\label{C.2}
H_{2} = 1 + P_2 z + \Bigl( \frac14 P_1 P_2 + \frac14 P_2 P_3 \Bigr) z^{2}
  + \Bigl( \frac{1}{36} P_2^{2} P_3 + \frac19 P_1 P_2 P_3 \Bigr) z^{3}
   + \frac{7}{288} P_1 P_2^{2} P_3 z^{4}  \\
   \nonumber
   + \Bigl( \frac{1}{576} P_1 P_2^{3} P_3 + \frac{1}{900} P_1^{2} P_2^{2} P_3 \Bigr) z^{5} +
   \Bigl( \frac{1}{6400} P_1^{2} P_2^{3} P_3 + \frac{1}{20736} P_1 P_2^{3} P_3^{2} \Bigr) z^{6}
    \\ \nonumber
    + \frac{1}{129600} P_1^{2} P_2^{3} P_3^{2} z^{7} + \frac{1}{8294400} P_1^{2} P_2^{4} P_3^{2} z^{8},\\
\label{C.3}
H_{3} = 1 + P_3 z + \frac12 P_2 P_3 z^{2} + \Bigl( \frac{1}{18} P_1 P_2 P_3 + \frac19 P_2^{2} P_3 \Bigr) z^{3}
 + \Bigl( \frac{1}{144} P_2^{2} P_3^{2} + \frac{1}{32} P_1 P_2^{2} P_3 \Bigr) z^{4}
 \\ \nonumber
+ \Bigl( \frac{1}{288} P_1 P_2^{2} P_3^{2} + \frac{1}{400} P_1^{2} P_2^{2} P_3 \Bigr) z^{5}
 + \Bigl( \frac{1}{2025} P_1^{2} P_2^{2} P_3^{2} + \frac{1}{10368} P_1 P_2^{3} P_3^{2} \Bigr) z^{6}
\\ \nonumber + \frac{1}{28800} P_1^{2} P_2^{3} P_3^{2} z^{7}
+ \frac{1}{921600} P_1^{2} P_2^{4} P_3^{2} z^{8} + \frac{1}{74649600} P_1^{2} P_2^{4} P_3^{3} z^{9}.
\ear


\subsection{The dimensionless variation of $G$}

Here we give a derivation of the relation (\ref{Appro}) for an
approximate value of the dimensionless parameter of relative
variation of $G$.

We start with the relation (\ref{varGd}) written in the following
form
  \beq{A.1}
       \delta = \dot{G}/(GH) = \frac{t^2- t^2_0}{|A P| t_0^2 t^2},
   \eeq
where  $H = \dot{a}_3/a_3$ is the Hubble parameter and
 $t_{0}$ is defined in  (\ref{extr}). (We recall
 that here and in what follows  $\dot{f} = df/d \tau$.)
 In the vicinity of the point
 $t_0$ we get in linear approximation
  \beq{A.2}
       \delta \approx  \frac{\triangle t}{|A P| t_0^3}.
   \eeq
Using the synchronous time variable  $\tau = \tau(t)$ we get
    \beq{A.2a}
   \delta \approx  \left(\frac{dt}{d \tau}\right)_0 \frac{\triangle \tau}{|A P| t_0^3}=
    \frac{\triangle \tau}{X^{A}_{0} |A P| t_{0}^{3}}
   \eeq
 ($d\tau/d t = X^{A}$). Here the subscript "0" refers to
 $t_0$. From (\ref{scf}) we can find $\dot{a}_{3}$
 \bear{adot}
 \dot{a_{3}} = \frac{dt}{d\tau}\frac{da_{3}}{dt} = \frac{2|A||P|t}{X}
 \ear
 and thereby for the Hubble parameter we get
  \beq{A.3}
    H_0 = \left(\frac{\dot{a}_3}{a_3}\right)_0 =
          \frac{2|A||P| t_0}{X^{A+1}_0}.
  \eeq
  Then, it follows from (\ref{A.2a}) and   (\ref{A.3}) that
  \beq{A.4}
   \delta \approx   \frac{X_{0} }{2 A^2 P^2 t_0^4} H_0 \triangle \tau.
  \eeq
  Since (see (\ref{funcH}) and (\ref{extr}))
 \beq{A.5}
  X_{0} = \frac{4 |A|}{1 +4 |A|}
 \eeq
 the pre-factor in (\ref{A.4}) reads (see (\ref{extr})):
  \beq{A.6}
   \Pi = \frac{X_{0} }{2 A^2 P^2 t_0^4} = 8 + 2 |A|^{-1}
  \eeq
 and  we are led to the relation
 \beq{A.7}
         \delta  \approx  \Pi  H_0 (\tau - \tau_0),
 \eeq
 coinciding with (\ref{Appro}).

 \begin{center}
 {\bf Acknowledgments}
 \end{center}
I am very grateful to V.D. Ivashchuk for helpful remarks.
This work was supported in part by the Russian Foundation for
Basic Research grant 09-02-00677-a and by a grant of People Friendship University of Russia (NPK MU).

{\small

 \end{document}